\documentclass[prl,amsmath,amssymb,aps,reprint,a4paper,superscriptaddress,citeautoscript,floatfix]{revtex4-2}

\usepackage{amsmath,amsfonts,amssymb}
\usepackage[pdftex]{graphicx}
\usepackage[svgnames]{xcolor}
\usepackage[colorlinks=True, linkcolor=DarkRed, citecolor=ForestGreen, urlcolor=MediumBlue, pdfstartview=FitH, bookmarks=False, pdfpagemode=UseNone]{hyperref}
\usepackage{bm}
\usepackage{color}

\newcommand{\vecr}{{\bm r}}
\newcommand{\bbar}{\overline{B}}
\newcommand{\ak}{a_{\bm k}}

\newcommand{\bk}{b_{\bm k}}
\newcommand{\kxy}{{\bm k}_{xy}}
\newcommand{\kz}{k_z}
\newcommand{\rxy}{\vecr_{xy}}

\newcommand{\cosxy}{\cos(\kxy\rxy)}
\newcommand{\sinxy}{\sin(\kxy\rxy)}
\newcommand{\cosz}{\cos(\kz z)}
\newcommand{\sinz}{\sin(\kz z)}

\begin{document}

\title{Realistic Pearl vortices in thin film superconductors}

\author{Aurélien Balzli}
\affiliation{Department of Quantum Matter Physics, University of Geneva, 24 quai Ernest-Ansermet, 1211 Geneva, Switzerland}
\author{Louk Rademaker}
\affiliation{Department of Quantum Matter Physics, University of Geneva, 24 quai Ernest-Ansermet, 1211 Geneva, Switzerland}
\affiliation{Institute-Lorentz for Theoretical Physics, Leiden University, PO Box 9506, 2300 Leiden, The Netherlands}
\author{Giulia Venditti}
\affiliation{Department of Quantum Matter Physics, University of Geneva, 24 quai Ernest-Ansermet, 1211 Geneva, Switzerland}

\date{\today}

\begin{abstract}

We analyze magnetic field profiles of vortices in thin-film superconductors, shedding new light on this old and presumed settled problem. 
In sufficiently thin films with realistic Ginzburg-Landau parameter $\kappa = 1/\sqrt{2}$, the magnetic screening around a vortex core is neither exponential - as is expected in bulk - nor the power-law that was predicted by Pearl. Instead, a universal curve for the magnetic field variation appears that scales with the sample thickness. The thickness dependence is consistent with the seminal Pearl length, and serves as an indication of the reduced magnetic field screening present in two-dimensional superconductor. Finally, we quantify the crossover from bulk-like to thin superconductors, and establish different screening length-scales relevant for the analysis of experimental data. 
\end{abstract}

\maketitle

{\em Introduction} -- The hallmark of a superconductor is its ability to screen an external magnetic field, characterized by an exponential decay of the field strength over the London penetration depth $\lambda$ \cite{tinkham_introduction_2004}. In type II superconductors, where the Ginzburg-Landau (GL) parameter $\kappa = \lambda / \xi$ exceeds the critical value $\kappa_c = 1/\sqrt{2}$, magnetic field penetration can take the form of a triangular Abrikosov vortex lattice \cite{Abrikosov1957, Hess1989, Volodin2000, Blatter1994, Varlamov2018}. The size of the vortex core is set by the correlation length $\xi$, and the field is exponentially suppressed away from the vortex core.

\begin{figure}
\centering
    \includegraphics[width=1\linewidth]{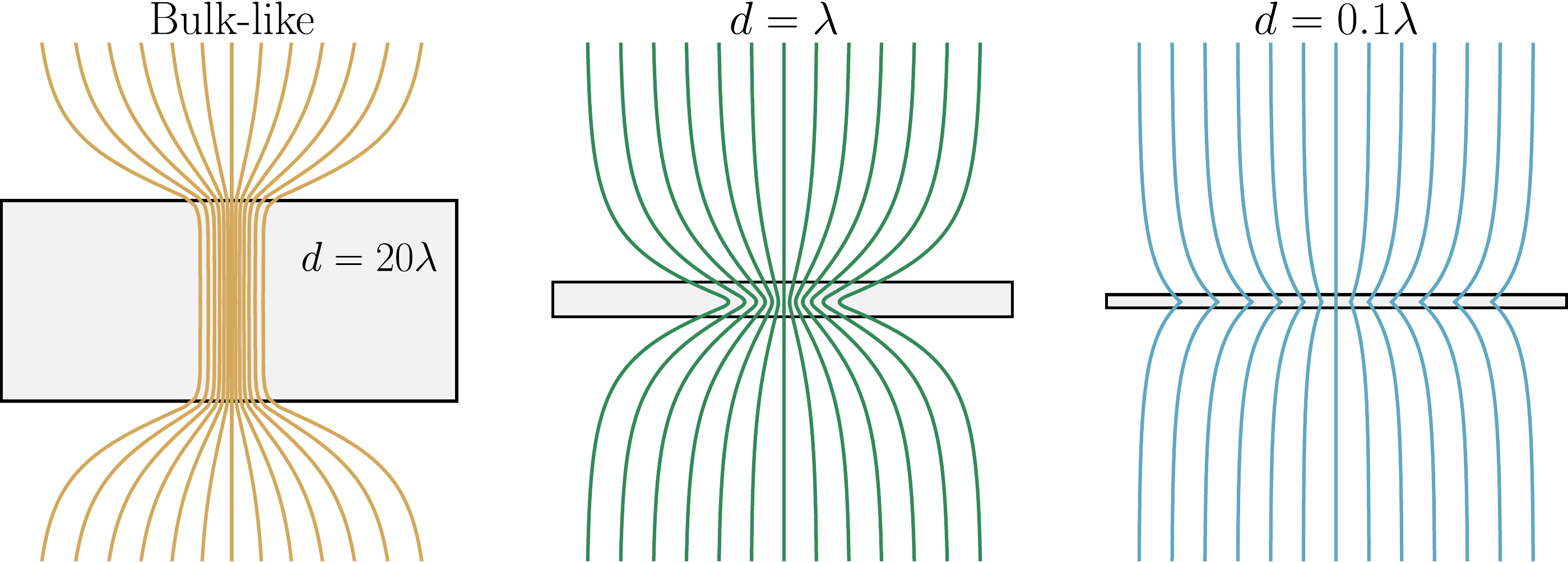}
    \caption{
Magnetic field profiles for different sample thicknesses $d$ for $\kappa=1/\sqrt{2}$ in the bulk ($d=20\lambda$), crossover dimension ($d=\lambda$) and thin-films ($d=0.1\lambda$).
}
	\label{fig:Cartoon}
\end{figure}

In thin-film superconductors, however, magnetic screening is no longer governed by the bulk penetration depth and heavily depends on the film thickness $d$, see Fig.~\ref{fig:Cartoon}. 
It was suggested that in this case the much larger Pearl length $\Lambda=2\lambda^2/d$ \cite{Pearl1964} determines the spatial extent of magnetic field screening, which is central to understanding superconducting devices. 
Moreover, the reduced screening in 2d superconductors is an essential ingredient in the description and observation of Berezinski-Kosterlitz-Thouless (BKT) physics in thin superconductors \cite{Beasley1979, Halperin1979, Kogan2007}, as it is supposed to set the scale below which the vortex-antivortex interaction is logarithmic. Since $\Lambda$ often exceeds the sample dimensions by orders of magnitude, it allows two-dimensional superconductors to develop the quasi-long-range order predicted by the BKT theory.
Despite its fundamental role, direct probes of the Pearl length remain challenging, motivating a renewed investigation into the electrodynamic response of superconducting thin films across length scales \cite{Tafuri2004,Keren2022, Kogan2021}.

The deep physical meaning of the Pearl length is often overlooked in literature. 
Specifically, Pearl estimated a different functional behavior of the vortex profiles compared to bulk vortices: in thin films, the magnetic field decays as $1/r$ near the vortex core and crosses over to a $1/r^3$ decay at large distances $r\gg\Lambda$ \cite{Pearl1964, Kogan2001}. 
This description relies on London theory, which assumes a constant superfluid density and a point-like vortex core ($\xi\xrightarrow{} 0$) and is therefore only valid in the extreme type-II limit of large GL parameters $\kappa\gg1$. 
For lower, realistic, values of $\kappa$, vortex core sizes are non-negligible, which can alter the magnetic field profile of thin film vortices.
Finite-core effects have been studied in previous works, but often restricted to bulk superconductors rather than thin films \cite{Ihle1971, Clem1975, Pogosov2000, Pogosov2001}.
It therefore remains unclear whether the Pearl description holds for real superconductors with nonzero vortex cores. 
In particular, materials with $\kappa\gtrsim1/\sqrt{2}$ such as 
Nb ($\kappa \approx0.85$)  \cite{McConville1965}, 
V ($\kappa \approx0.7-0.9$) \cite{Moser1982,vanGurp1967}, 
LaPtSi$_3$ ($\kappa \approx2.49$) \cite{Smidman2014}, 
and the transition metal pnictide WP ($\kappa \approx1.3$) \cite{Nigro2022} are not expected to fit Pearl's description.
Additionally, thin films usually exhibit dirty superconductivity, leading to an increased penetration depth \cite{Gubin2005,Lemberger2007}, and consequently to an increased $\kappa$ compared to their bulk counterparts. 
This implies that thin films of type-I superconductors can exhibit mixed state configurations \cite{Gladilin2015,Cadorim2019, brandt_ginzburg-landau_2005, Maki1965, Palonen2013, Lasher1967}.
Indeed, a clear distinction between type-I and II superconductors breaks when dealing with thin films, and $\kappa_c$ itself depends on the thickness in a non-trivial way \cite{Cadorim2019, Crdoba-Camacho2016}.
Nevertheless, experimental observations of Pearl vortices yield at least qualitative agreement with the theoretical prediction 
\cite{Tafuri2004, hao_direct_2006, Keren2022, fridman_anomalous_2025}.
The formation of giant vortices in thin films consistent with a Pearl description, as well as their phase diagram, and 
vortex-vortex interactions \cite{Lasher1967, Callaway1992, Schweigert1998, Shi2009, Berdiyorov2006,Palonen2013, Crdoba-Camacho2016} were already investigated, although the functional dependence of such vortices was not properly addressed.

In this Letter, we study the magnetic field profile for realistic values of $\kappa$, in an {\em intermediate} regime between putative type I and type II superconductors, for a broad range of thicknesses $d$. 
To do so, we perform exact numerical simulations of dilute vortex lattices using GL theory. 
Our analysis of the magnetic screening length demonstrates a thickness dependence different from the one proposed by Pearl, as well as a different functional form of magnetic profiles.
Specifically, we observe large vortex-core effects, in contrast with Pearl's assumption of a point-like core.
Despite these functional changes, we confirm the relevance of the Pearl length: for sufficiently thin films, it sets the overall scale of the magnetic field profile. 


{\em Methods} -- To find the magnetic field surrounding a vortex in thin film superconductors of thickness $d$, we consider an infinite vortex lattice in the $x$--$y$ plane, where each unit cell carries a single quantum of flux $\Phi_0$. The average magnetic field $\overline{B}$ is thus $\Phi_0$ divided by the unit cell area $S$.
Given a triangular lattice of 
 vortices, the vortex positions can be expressed as $\vecr_{xy}\equiv \vecr_{xy}^{mn} =  a(m + n/2, n\sqrt{3}/2)$, where $a$ is the inter-vortex distance. 
In the superconducting phase, the GL free energy density expressed in reduced units takes the form \cite{brandt_ginzburg-landau_2005, brandt_precision_1997}
\begin{equation}
    \begin{split}
    f =  \Bigl \langle  - & |\psi(\vecr)|^2 + \frac{|\psi(\vecr)|^4}{2} +\\
    & + \left|\left(\frac{\nabla}{i\kappa} - {\bm A}(\vecr)\right) \psi(\vecr)\right|^2
    + B(\vecr)^2 \Bigr \rangle
    \end{split}
\end{equation}
where the free energy is measured in units of $B_{c1}^2/4\pi$, the magnetic field in $\sqrt2B_{c1}$, $B_{c1}$ being the lower critical field, and lengths in units of the London length $\lambda$.
Here $\psi(\vecr)$ is the complex superconducting order parameter, and ${\bf A}(\vecr)$ the vector potential of the local magnetic field.
Angle brackets represent spatial averaging. 
In films of finite thickness, one must consider the stray field outside the sample. 

Defining the superfluid density as $\rho(\vecr)=|\psi(\vecr)|^2$,  
and the {supercurrent} as ${\bm Q}(\vecr) = \left(\frac{\nabla}{i\kappa}-{\bm A}(\vecr)\right)\psi(\vecr)$,
such that the magnetic field is ${\bm B}(\vecr)=\nabla\times{\bm Q}(\vecr)$, we can rewrite the free energy density of a thin film as follows
\begin{equation}
    \begin{split}
     f =  \Bigl \langle & -\rho(\vecr) +\frac{\rho(\vecr)^2}{2} + g(\vecr) + \\ 
    & + \rho(\vecr){\bm Q}(\vecr)^2 + (\nabla\times{\bm Q}(\vecr))^2 \Bigr \rangle + \frac{F_{\mathrm{stray}}}{d}, \\
\end{split}
\end{equation}
where the stray field free energy is 
\begin{equation}
        F_\mathrm{stray} = 2\int_{d/2}^\infty \langle {\bm B}(\vecr)^2-\overline B^2\rangle_{x,y}\ dz, 
\end{equation}
with $g(\vecr)=\frac{(\nabla\rho(\vecr))^2}{4\kappa^2\rho(\vecr)}$, provided that the superconductor occupies the region $-d/2<z<d/2$.
Using the periodicity of the problem, we express the superfluid density and supercurrent using a Fourier expansion,
\begin{align}
    & \rho(\vecr) = \sum_{\bm k} \ak(1-\cosxy)\cosz\\
    &{\bm Q}(\vecr) = {\bm Q}_A(\rxy) + \sum_{\bm k}\bk\frac{\hat{z}\times\kxy}{\kxy^2}\sinxy\sinz.
\end{align}
Here ${\bm k}=({\bm k}_{xy},\text{k}_z)$ and the ${\bm k}_{xy}=2\pi/\vecr_{xy}$ are the reciprocal vectors of the vortex lattice. The expansion ensures a vanishing order parameter at vortex positions $\vecr_{xy}$. 
The supercurrent is expanded in Fourier modes around the Abrikosov solution for the vortex lattice ${\bm Q}_A$, valid at the upper critical field $B_{c2}$. 

The minimization of the free energy leads to a self-consistent formulation 
for the Fourier coefficients $\ak$, $\bk$. 
Taking the Abrikosov lattice $\rho_A$, ${\bm Q}_A$ as an initial solution, the iteration procedure detailed in \cite{brandt_ginzburg-landau_2005} 
allows us to find solutions for any value of an applied field $b\equiv \overline B/B_{c2}$, GL parameter $\kappa$, and thickness $d$.
We can then compute the magnetic field:
\begin{align}
    & {\bm B}(\vecr) =\bbar\hat{z} + {\bm b}(\vecr), \quad \langle{\bm b}(\vecr)\rangle=0\\
    & {\bm b}_z(\vecr) = \sum_{\bm k}\bk\cosxy\cosz\\
    & {\bm b}_{xy}(\vecr) = \sum_{\bm k}\bk\frac{\kxy \kz}{\kxy^2}\sinxy\cosz.
\end{align}
Outside the sample, the field obeys the Laplace equation in vacuum and each coefficient decays with $e^{-k_{xy}(z-d/2)}$.


\begin{figure}
	\centering
	\includegraphics[width=\linewidth]{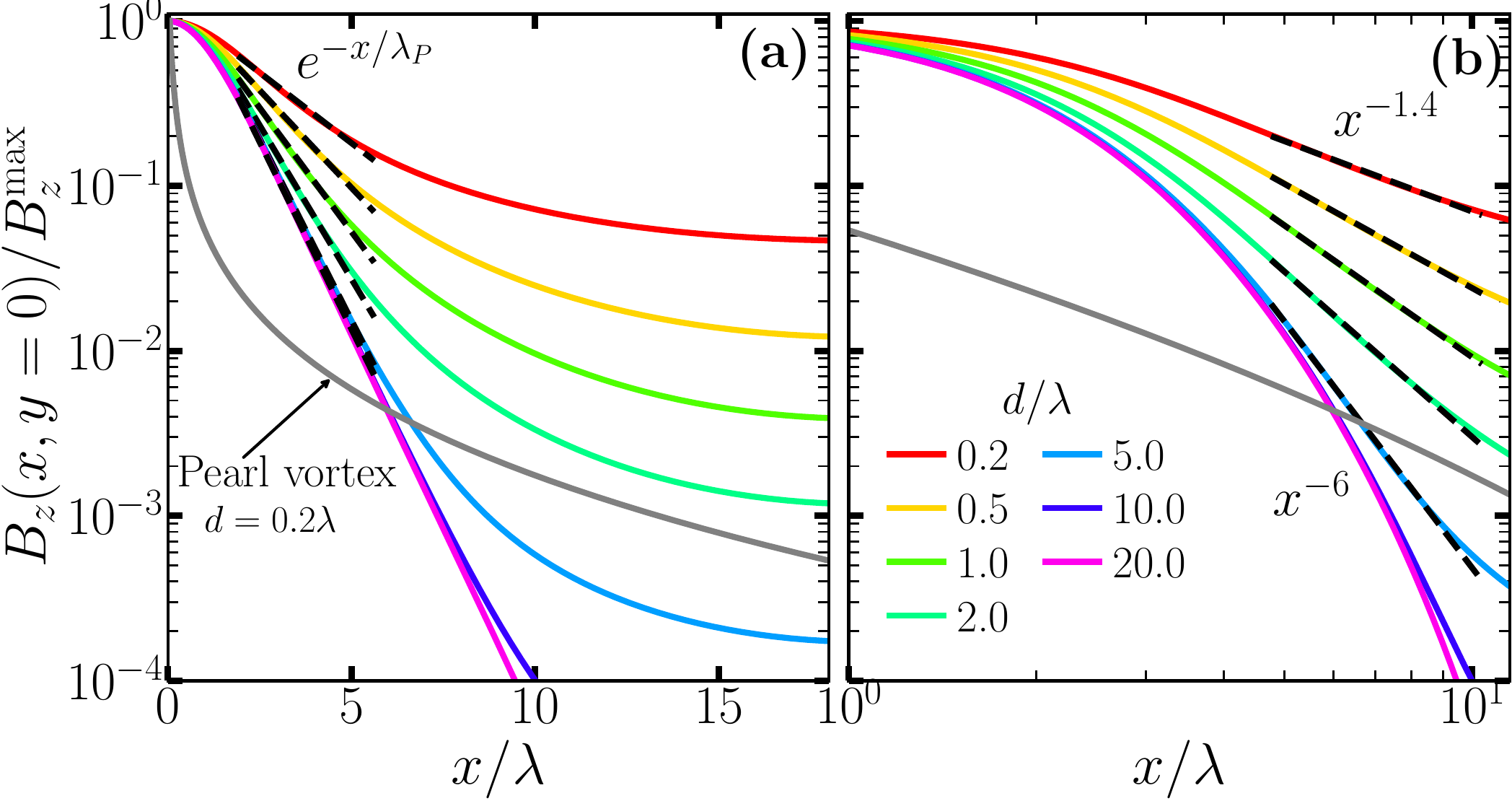}
	\caption{
        Magnetic field profiles in the center of films of different thicknesses $d$ in \textbf{(a)} lin-log and \textbf{(b)} log-log scales. 
        The normalized magnetic fields $B_z/B_z^\text{max}$ are plotted against the distance from the center of the vortex in the $x$ direction. Length scales are in units of $\lambda$. 
        We do not observe the expected Pearl behavior $1/r$ near the vortex core and $1/r^3$ far from it.
        Gray curves are the field profiles of a Pearl vortex, computed from the analytical expression derived in \cite{Carneiro2000}.
        }
	\label{fig:Profiles}
	\end{figure}

{\em Magnetic field profiles} -- 
To analyze the magnetic field profiles in the intermediate regime between type I and type II superconductors, we focus on our results for $\kappa=1/\sqrt{2}$. 
We consider a low field $b=0.01$ to simulate a low vortex density, to obtain a relatively small overlap between neighboring vortices. 
Magnetic field profiles in the center of films for various film thicknesses are shown in Fig.~\ref{fig:Profiles}, respectively in lin-log and log-log scales in panels \textbf{(a)} and {\bf (b)}.
In a thick film with $d=20\lambda$, we recover the expected exponential screening of the magnetic field [see the magenta curve in  Fig~\ref{fig:Profiles}\textbf{(a)}]. 
Field profiles of thicker samples overlap with the $d=20\lambda$ curve, proving that the system reached its bulk limit.
Thinner films instead show a broadened field profile, indicating increased field penetration, or equivalently, reduced screening. 
While this is in qualitative agreement with Pearl's description, there is a significant difference between our results and the functional behavior expected from Pearl's solution.
In particular, we do not observe the expected $1/r$ or $1/r^3$ power-law behaviors anywhere, which we plot with gray curves for comparison.
Near the vortex core, we find an exponential decay [Fig.~\ref{fig:Profiles}\textbf{(a)}], while, at larger distances, we identify a range of power-laws {with thickness-dependent exponents} [Fig.~\ref{fig:Profiles}\textbf{(b)}].
The smallest thickness plotted is $d=0.2\lambda$, which can already be considered as a very thin film, as we will show below; $0.2\lambda< d\leq 10\lambda$ is a crossover region between thick and thin films. 


{\em Magnetic lengths} -- 
While the magnetic field profiles do not correspond to the Pearl picture, we can, however, extract relevant magnetic length scales that characterizes the reduced screening in thin films. 

\begin{figure}
    \centering
    \includegraphics[width=1\linewidth]{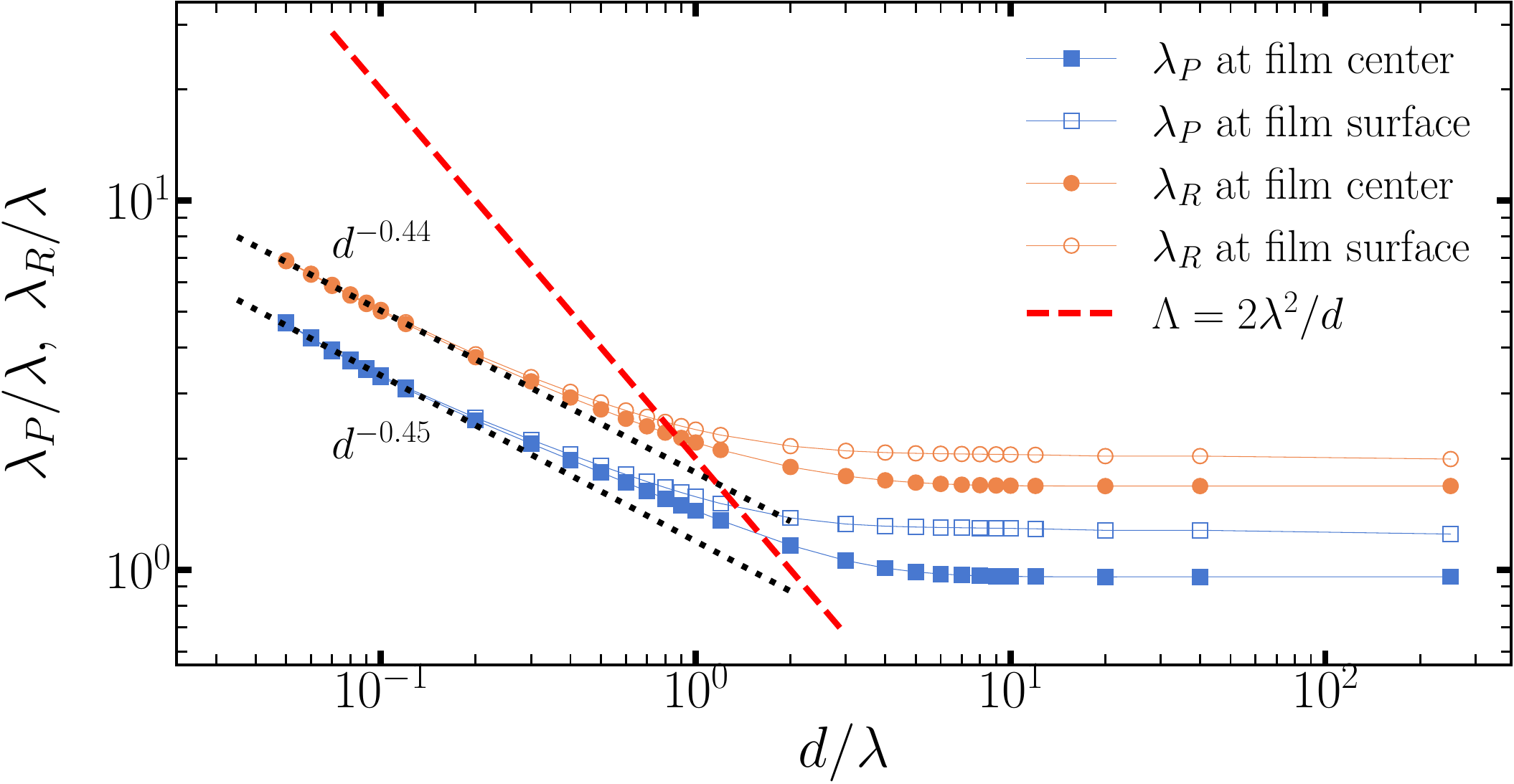}
    \caption{
    Magnetic screening lengths. 
     The values of $\lambda_{P}$ (circles) and $\lambda_{R}$ (squares) are computed both at the film center {(full symbols)} and surface (empty symbols). They exhibit a power-law scaling in very thin films, where the center and surface values meet. 
     The dashed red line is the Pearl length.}
    \label{fig:Magnetic_lengthscales}
\end{figure}

One possible way to characterize the reduced screening is through a generalized penetration depth,
\begin{align}
    \lambda_R \equiv B_z(0)^{-1}\int_0^\infty B_z(r) dr,
\end{align}
which reduces to the bulk London penetration depth $\lambda$ in the case of exponential screening, $B_z(r)=B_z(0)e^{-r/\lambda}$. 
Its thickness dependence is shown in orange in Fig.~\ref{fig:Magnetic_lengthscales} at the surface (empty circles) and at the center (full circles) of the film.  
For systems thinner than $d\lesssim0.2\lambda$, the generalized penetration depth is the same in the center as on the surface of the film, and therefore at these thicknesses we have reached the two-dimensional limit. 
While the increase of $\lambda_R$ clearly signals reduced screening, it does not follow the Pearl length description. Rather, we can fit the length with a power law $\lambda_R \sim d^{-0.44}$, as shown with dotted lines in Fig.~\ref{fig:Magnetic_lengthscales}.
Same results are obtained for other values of $\kappa$.\cite{Supplementary}

An alternative definition of the magnetic screening length is to fit the magnetic field profile near the vortex core with an exponential, as is shown in Fig.~\ref{fig:Profiles}{\bf (a)}. This length $\lambda_P$ approaches the bulk London penetration depth for thicknesses $d\gg \lambda$.
Its thickness-dependence is plotted with blue squares in Fig.~\ref{fig:Magnetic_lengthscales}, either at the film center (full symbols) and at the surface (empty symbols).
The two length scales $\lambda_P$, $\lambda_R$ show the same functional dependence, both in the surface and at the center -- which is fundamentally different than the behavior of the Pearl length.

These results remain valid at least up to $\kappa=2$. Above $\kappa=2$, the functional behavior of $\lambda_P$ starts to deviate from $\lambda_R$. We attribute this effect to the large overlap between vortices, which prevents us from studying the properties of isolated vortices. 
See the Supplementary Material \cite{Supplementary} for additional information and data.

{\em Reviving the Pearl length --}
\begin{figure}
    \centering
    \includegraphics[width=1\linewidth]{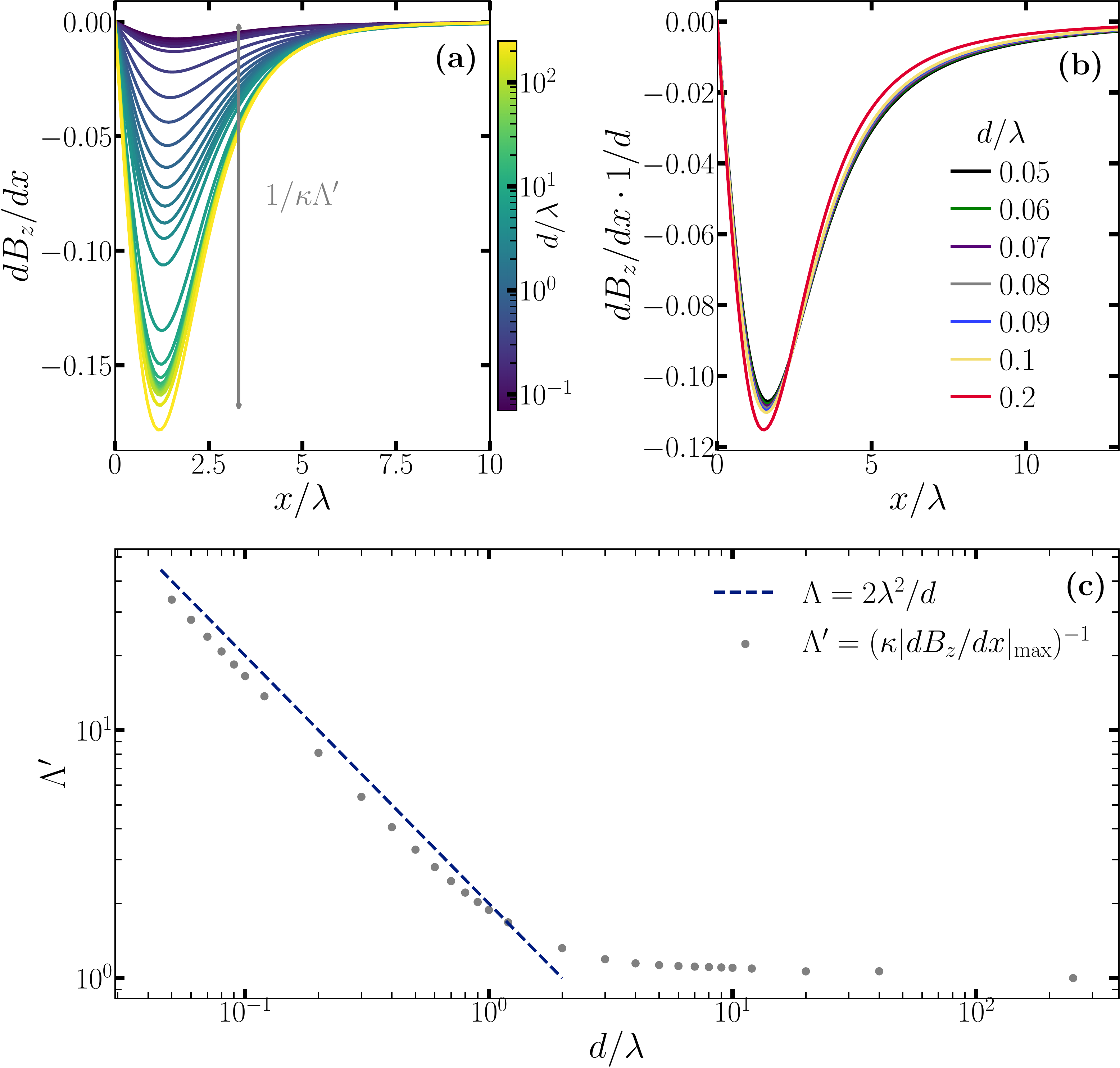}
    \caption{\textbf{(a)} Derivative of the magnetic field at the film surface for various thicknesses $d$ as a function of the inter-vortex distance $x$. 
     \textbf{(b)} Rescaling them with respect to their thickness, they all fall on the same curve, confirming the $1/d$ behavior predicted by Pearl, but with a different functional form.
    \textbf{(c)} The length $\Lambda'$ extracted from the field derivative amplitudes shown in the top panel, following Eqs. \eqref{Eq:NewPearlLength}. Here $\Lambda'$ is normalized by its value for $d=250\lambda$. 
    }
    \label{fig:Exp_comparison}
\end{figure}
The magnetic field profiles we find do not follow the power-law description of Pearl, nor do the length scales $\lambda_{R,P}$ we extract follow Pearl's thickness dependence. However, we do find a new relevance for the Pearl length $\Lambda$ in that it quantifies the {\em strength of magnetic field variations}. 

Compared to a vortex in a bulk superconductor, the magnetic field profile in thin films is much smoother. This can be quantified by considering the field derivative $dB_z/dx$, as shown in Fig.~\ref{fig:Exp_comparison}{\bf (a)} at the surface of the thin film. Surprisingly, we find that for films in the two-dimensional limit where $d \lesssim 0.2 \lambda$ the magnetic field derivative is a universal curve multiplied by the film thickness. Therefore, when we plot $dB_z/dx$ \textit{divided} by the thickness $d$, all curves collapse onto each other, see Fig.~\ref{fig:Exp_comparison}{\bf (b)}. 

The inverse of the maximum field derivative thus defines a new length scale
\begin{align}
    \Lambda' = \left( \kappa  \left|\frac{dB_z}{dx}\right|_{\rm max} \right)^{-1}
    \label{Eq:NewPearlLength}
\end{align}
which, as we show in Fig.~\ref{fig:Exp_comparison}{\bf (c)}, follows closely the original Pearl length $\Lambda = 2 \lambda^2/d$. In fact, the strength of the magnetic field variation in the picture of Pearl is also proportional to the film thickness. While the magnetic profile is different, the thickness dependence is the same in our results as in Pearl's,
\begin{align}
    \label{eq:Pearl}
    \text{Pearl:}\qquad & \frac{dB_z(r)}{dr} = -\frac{\Phi_0}{4 \pi \lambda^2 r^3} d =  - \frac{\Phi_0}{2\pi \Lambda r^3}\\
    \label{eq:BRV}
    \text{This work:}\qquad & \frac{dB_z(r)}{dr}= f(r) d \propto \frac{f(r)}{\Lambda'}
\end{align}
with $f(r)$ the function shown in Fig.~\ref{fig:Exp_comparison}{\bf (b)}. The origin of this universal function stems from the fact that in the two-dimensional limit, the screening supercurrents are independent of the vertical coordinate $z$. As a result, the total amount of screening currents is proportional to the thickness of the film.


{\em Comparison to experiments --} We can compare our results to a recent experimental measurement of the magnetic field profile around a vortex in thin film NbSe$_2$ \cite{Kogan2001,fridman_anomalous_2025}. There, SQUID-on-tip microscopy measurements was used to determine the magnetic field near the vortex cores. 
Based on Pearl's results, Eq.~\eqref{eq:Pearl}, the field derivative at a height $h$ above the surface would be
\begin{align}
    \label{eq:dBz}
    &\frac{\partial B_z(\vecr)}{\partial \vecr}=-\frac{\Phi_0}{2\pi\Lambda}\frac{ \text{r}}{[h^2+\vecr^2]^{3/2}}.
\end{align}
relative to the vortex position $\vecr_j$. Since this field derivative is inversely proportional to the Pearl length, $\Lambda$ was then extracted from the inverse of the maximum of the experimentally measured $dB_z/dx$ curves, similar to our Eq.~\eqref{Eq:NewPearlLength}. 

\begin{figure}
    \centering
    \includegraphics[width=\linewidth]{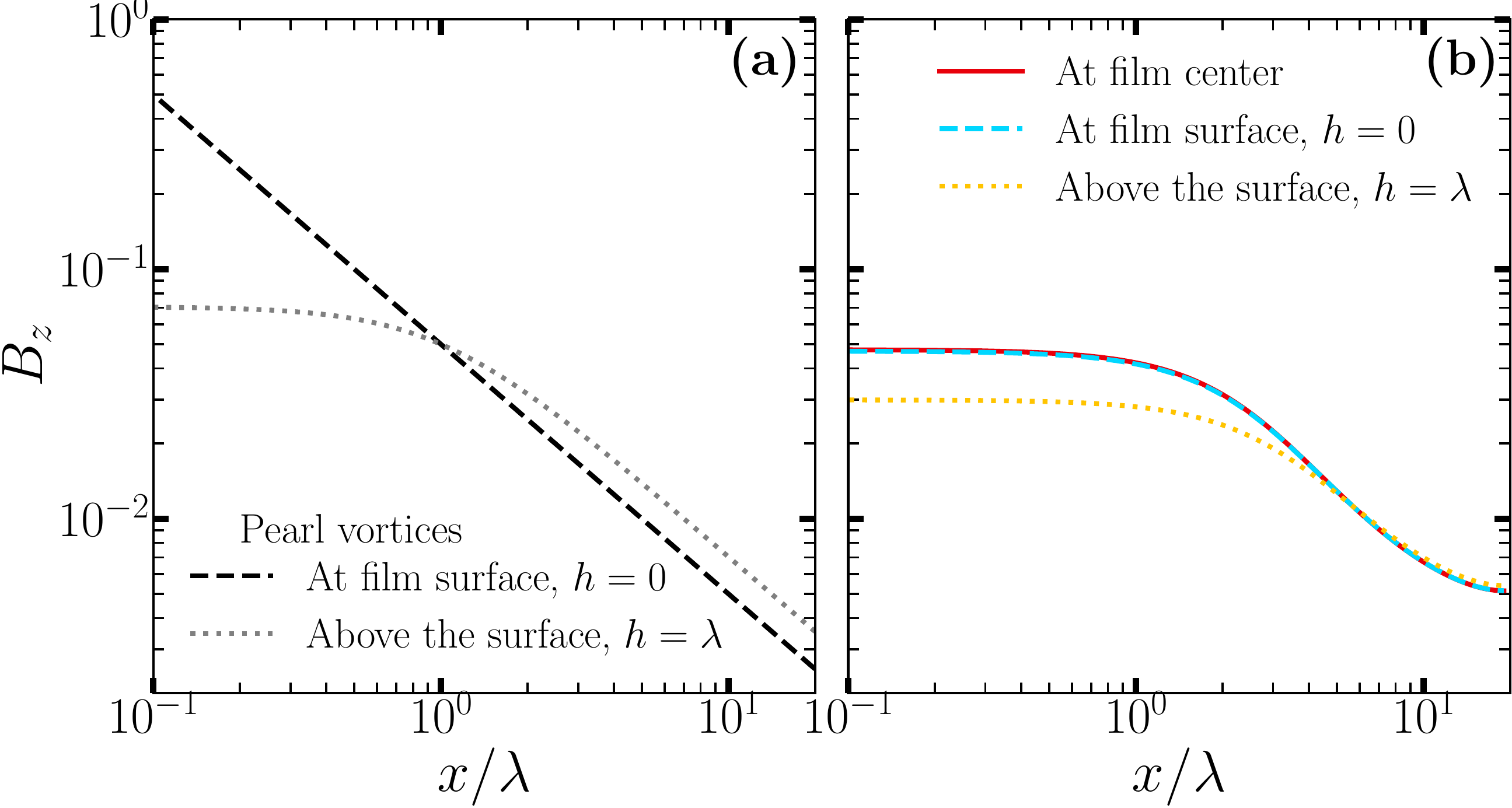}
    \caption{
    \textbf{(a)} Magnetic field profiles expected from the Pearl's description of vortices. See Eqs (\ref{eq:Pearl}) and (\ref{eq:dBz}) in the main text. 
    \textbf{(b)} Magnetic field profiles at different heights within a film of thickness $d=0.1\lambda$. While at the surface the exact profile is much different from the Pearl prediction, at a finite height above the thin film the difference between Pearl and our results is less clear.
    }
    \label{fig:Fields_vs_Pearl}
\end{figure}

As shown in Fig.~\ref{fig:Fields_vs_Pearl}, at a finite height above the thin film, the Pearl field profile following Eq.~\eqref{eq:Pearl} and \eqref{eq:dBz} is difficult to distinguish qualitatively from our exact numerical results. There is a fundamental difference, however: In our results the inflection point of the magnetic field profile is an intrinsic property of the thin film, whereas in the Pearl result it only appears because the field profile is analyzed at an height $h$ above the surface. A precise measurement of the field profile dependence on the height $h$ allows to distinguish between Pearl and our result.

{\em Outlook} --
In conclusion, we computed magnetic profiles of vortices in thin film superconductors for realistic values of the GL parameter, focusing on $\kappa = 1/\sqrt{2}$. We find a different magnetic field profile than the power-law decay predicted by Pearl. However, the Pearl length still persists as the relevant length scale for screening in thin films. In the two-dimensional limit, $d \lesssim 0.2\lambda$, the magnetic field profile is a universal function that scales with the film thickness. Our new Pearl length, based on the derivative of the magnetic field using Eq.~\eqref{Eq:NewPearlLength}, closely follows the `original' Pearl length. 
The transition from the pure London exponential screening to this unusual longer-ranged Pearl-like screening happens in a regime of thicknesses $0.2\lambda< d\leq 10\lambda$.

We verified that the results presented here remain valid at least up to $\kappa=2$. Beyond that value, and within the numerical method presented here, the analysis of an isolated vortex become subject to overlap with neighboring vortices.
Further work is needed to investigate at which value of $\kappa$ the vortex size becomes negligible and Pearl solution applies.
Extending variational approaches~\cite{Clem1975} or the circular cell method \cite{Ihle1971,Pogosov2000,Pogosov2001} to thin films may allow for such a study.

The values of $\kappa$ that we used are the bulk values, which are constant and well defined, whereas thinner films display a thickness-dependent penetration depth. The critical value $\kappa_c$ setting the extent of the mixed state is itself thickness dependent \cite{Cadorim2019}. Therefore, an experimentally reported increase of a screening length with decreasing film thickness should be cautiously analyzed taking into account a possible shift of $\kappa$. 

With the discovery of atomically thin superconductors in graphene and TMD moiré structures
(see eg.~\cite{Zhang2026,Xia.2026}), there exist now two-dimensional superconductors who do not have a bulk equivalent. It is an interesting open question to what extent the manifestly three-dimensional Ginzburg-Landau free energy can be applied to the study of screening currents in atomically thin superconductors.

Our quantitative results show that while the Pearl length is still the relevant length scale to describe the screening in thin film superconductors, the shape of the magnetic field profile can be drastically different. With advances in the detection of magnetic fields on the nanometer scale, it will be very interesting to see an experimental verification of the correct field profile of thin film superconductors.

{\em Acknowledgments ---} 
We thank Ilaria Maccari and Christophe Berthod for fruitful discussions. 
G.V. acknowledges financial support from the Swiss National Science Foundation (SNSF) via Swiss Postdoctoral Fellowship TMPFP2 224637. 
L.R. acknowledges the Swiss National Science Foundation (SNSF) via Starting Grant TMSGI2 211296.

\bibliography{biblio}

\end{document}